\documentclass{article}

\usepackage{spconf,hyperref}
\usepackage{cite}
\usepackage{amsmath,amssymb,amsfonts}
\usepackage{algorithmic}
\usepackage{graphicx}
\usepackage{subcaption}
\usepackage{textcomp}
\usepackage{xcolor}
\usepackage{url}
\usepackage{cleveref}
\usepackage{booktabs}
\def\BibTeX{{\rm B\kern-.05em{\sc i\kern-.025em b}\kern-.08em
    T\kern-.1667em\lower.7ex\hbox{E}\kern-.125emX}}
\begin{document}

\title{Look Less, Hear Better: Jointly Rewarded GRPO for Streaming ASR}
%
\name{Xiuwen Zheng}
\address{University of Illinois Urbana-Champaign}

\maketitle

\begin{abstract}
Streaming automatic speech recognition (ASR) must be judged jointly on what it
transcribes and on how quickly it commits each word. Delayed streams modeling
(DSM) has become the dominant paradigm for streaming large audio-language
models, exposing a structural delay $\tau$ that bounds the decoder's lookahead.
We show that $\tau$ is a poor proxy for user-perceived latency, and that the
alignment-based supervision of DSM leaves latency on the table: the same
forced-aligned transcript is used at every $\tau$, forcing the model to withhold
words it could already commit. We introduce AWED, a word-level emission-delay
metric defined relative to the acoustic end of each word, and post-train a DSM
recognizer with GRPO under a reward that scores transcription accuracy and
measured delay jointly. Trained at a single operating point ($\tau=6$ frames),
our model dominates both its supervised fine-tuning initialization and the
Voxtral Realtime backbone across all evaluated lookahead budgets: it cuts WER by
30.8\% relative at an 80\,ms structural delay, and by 5.7\% relative at 480\,ms
while lowering median AWED from 1.17\,s to 1.04\,s. Latency-rewarded
post-training thus advances the accuracy--latency Pareto frontier of streaming
ASR without architectural change.
\end{abstract}


\section{Introduction}
 
Streaming automatic speech recognition (ASR) underpins live captioning, voice
assistants, meeting transcription and simultaneous interpretation. In all of
these settings the quality a user perceives depends not only on whether words
are transcribed correctly, but also on \emph{when} they appear. Streaming ASR is
therefore inherently a two-objective problem, and progress is properly judged by
the accuracy--latency Pareto frontier rather than by word error rate (WER) alone.
 
Large audio-language models (LALMs) have recently achieved highly competitive performance on ASR benchmarks~\cite{liu2026voxtral}, and delayed streams modeling (DSM)~\cite{zeghidour2025streaming} has
emerged as the dominant recipe for making them streaming: text and audio are
placed on a shared time axis, with the transcript offset from the audio by a
structural delay $\tau$ that bounds the lookahead available to the decoder.
Voxtral Realtime~\cite{liu2026voxtral} builds on DSM with a causal streaming audio
encoder and conditions a single model on $\tau$ sampled from 1 to 30 frames
(80--2400\,ms), attaining state-of-the-art streaming accuracy across a broad
range of lookahead budgets. In parallel, reinforcement learning (RL)
post-training with group relative policy optimization
(GRPO)~\cite{shao2024deepseekmath} has proven effective for LLMs and, more recently, for speech
tasks~\cite{shivakumar2025grpo} --- but always under accuracy-only rewards.
 
Two gaps remain. \textbf{First, latency in DSM is declared rather than
measured.} The structural delay $\tau$ is a nominal bound on lookahead, not the
time a user waits for a word: at $\tau=1$ the median word-level emission delay of
Voxtral Realtime is 0.77\,s, nearly an order of magnitude above the nominal
80\,ms. The discrepancy arises because a word is committed only after its
aligned end time plus $\tau$, after any hesitation of the decoder, and after
enough subsequent tokens have been emitted to disambiguate it. Existing latency
measures --- token emission delay for RNN-T, average lagging and its
differentiable variants from simultaneous translation, or first-token latency ---
were not designed for the alignment-anchored output stream of DSM and do not
expose this gap. What is not measured cannot be optimized.
 
\textbf{Second, alignment-supervised training pins emission to a fixed
schedule.} Under DSM the training signal instructs the model to emit each word
exactly $\tau$ frames after its forced-aligned end time, and the same alignment
is reused for every $\tau$. When $\tau$ is large the decoder holds far more
acoustic evidence than its decision requires, yet the target forbids it from
committing earlier; the schedule is likewise uniform across words, ignoring that
many words are unambiguous well before their acoustic offset. This slack is out
of reach for supervised fine-tuning (SFT), whose objective is to imitate one
predetermined schedule, and equally out of reach for a hand-designed alignment
heuristic, which would require knowing the optimal per-word emission time in
advance. Emission timing is discrete and non-differentiable --- precisely the
regime where RL applies.
 
We bridge both gaps. We introduce AWED (aligned word emission delay), a
word-level latency metric defined relative to the acoustic end of each word and
reported over its full distribution, which makes the true cost of a given $\tau$
visible and differentiates systems that share the same nominal delay. We then
post-train a DSM streaming recognizer with GRPO under a reward that scores
transcription accuracy and measured emission delay jointly, gated by a match
rate that withholds latency credit from rollouts whose WER is too high for early
emission to be meaningful. Trained at a single operating point ($\tau=6$), the
resulting policy dominates both its SFT initialization and the Voxtral Realtime
backbone across the entire latency spectrum, reducing WER by 30.8\% relative at
the tightest 80\,ms budget at no latency cost.
 
Our contributions are as follows:
\begin{enumerate}
  \item AWED, a word-level emission-delay metric tailored to DSM, which
  separates true output latency from the nominal structural delay.
  \item A two-stage recipe that reproduces the Voxtral Realtime SFT pipeline and
  post-trains it with GRPO under a joint accuracy--latency reward.
  \item A consistent advance of the accuracy--latency Pareto frontier of
  streaming ASR, generalizing from a single training $\tau$ to all evaluated
  lookahead budgets.
\end{enumerate}
\section{Method}

\subsection{Delayed Streams Modeling and Supervised Baseline}

DSM places the audio and text streams on a shared time axis discretized at
$1/\Delta = 12.5$\,Hz ($\Delta = 80$\,ms). At each frame $t$ the decoder emits
one token $a_t \in \mathcal{V} \cup \{\texttt{[P]}\}$, where the pad symbol
$\texttt{[P]}$ denotes \emph{wait}; emission timing is therefore learned
end-to-end rather than governed by an external policy. The text stream is offset
from the audio stream by a structural delay of $\tau$ frames, which bounds the
lookahead available when any token is produced. Following~\cite{liu2026voxtral}, $\tau$
is injected into the decoder through AdaRMSNorm conditioning and sampled per
utterance during training, so that a single model serves every operating point
in the range at inference time.

Given a transcript $w_1,\dots,w_K$ with forced-aligned end times
$e_1,\dots,e_K$, the supervised target is constructed by placing the tokens of
$w_k$ contiguously from frame
\begin{equation}
  t_k \;=\; \lceil e_k / \Delta \rceil + \tau ,
  \label{eq:target}
\end{equation}
and filling every remaining frame with $\texttt{[P]}$. Equation~\eqref{eq:target}
is the object of interest for the rest of this paper: it prescribes a single
emission schedule, shifted rigidly by $\tau$, for all words and all operating
points. We additionally append $\tau_{\mathrm{pad}}$ pad frames after $e_K$
before the end-of-sequence symbol; anchoring EOS directly to the final word
causes the model to terminate prematurely and inflates deletion errors.

This stage reproduces the DSM recipe on our own aligned corpora. It is not a
contribution in itself, but it supplies both a like-for-like reference point and
the initialization $\pi_{\mathrm{ref}}$ for the post-training stage below.

\subsection{Aligned Word Emission Delay}
\label{sec:awed}

Latency in DSM is conventionally reported as $\tau$, which is a nominal bound on
lookahead rather than the time a user waits for a word. We therefore measure
emission delay directly, at word level, relative to the acoustic end of each
word.

Let a hypothesis consist of words $\hat{w}_1,\dots,\hat{w}_J$, and let
$f(j)$ be the frame at which the final token of $\hat{w}_j$ is emitted. We align
reference and hypothesis by word-level edit distance and retain the set
$\mathcal{M}$ of matched pairs $(k,j)$. For each such pair,
\begin{equation}
  \mathrm{AWED}(k,j) \;=\; f(j)\,\Delta \;-\; e_k ,
  \label{eq:awed}
\end{equation}
i.e.\ the audio duration consumed at the moment the word is committed, minus the
time at which the word actually ended in the signal. Delays are averaged within
an utterance and aggregated over a corpus by quantile; we report the median
(p50) as the primary figure and p90/p95 to expose the tail.

Restricting~\eqref{eq:awed} to $\mathcal{M}$ is not a convenience: a deleted
reference word has no emission time and an inserted hypothesis word has no
reference time, so emission delay is undefined outside the matched set. This
property is what makes an unconstrained latency objective exploitable, and
motivates the gate introduced in Sec.~\ref{sec:grpo}.

Anchoring to $e_k$ decouples the measurement from word duration and speaking
rate, which distinguishes AWED from existing measures. Token emission delay for
RNN-T is defined on that model's alignment lattice; average lagging and its
differentiable variants from simultaneous translation quantify read/write lag
against input \emph{length} rather than against an acoustic landmark; first-token
latency observes only the onset of the stream. None of the three exposes the gap
we are concerned with: at $\tau=1$, a nominal delay of $80$\,ms corresponds to a
median AWED of $0.77$\,s.

\subsection{Latency-Rewarded GRPO}
\label{sec:grpo}

Under DSM the state at frame $t$ is the conditioning delay together with the
consumed audio and the tokens emitted so far, and the action is the token
emitted next --- including $\texttt{[P]}$ and EOS. \emph{What} to emit and
\emph{when} to emit it thus inhabit the same discrete action space, and emission
timing can be optimized without any additional head or auxiliary decision model.
Because that timing is discrete and the delay in~\eqref{eq:awed} is a
non-differentiable function of the sampled trajectory, we optimize it by policy
gradient rather than by a modified supervised target.

For each utterance we draw a group of $G$ rollouts from the current policy at a
fixed conditioning delay and score rollout $i$ by
\begin{equation}
  r_i \;=\; \bigl(1 - \mathrm{WER}_i\bigr)
        \;-\; \lambda \, \mathbb{1}\!\left[m_i \ge m_0\right] \tilde{d}_i ,
  \label{eq:reward}
\end{equation}
where $m_i$ is the fraction of reference words matched by rollout $i$,
$\tilde{d}_i$ is the group-normalized emission delay
\textcolor{black}{$\tilde{d}_i = (d_i - \min_{i'} d_{i'}) / (\max_{i'} d_{i'} -
\min_{i'} d_{i'} + \epsilon)$ with $d_i$ the mean AWED of rollout $i$}, and $\lambda$ trades accuracy against latency.
Normalizing within the group places both terms on a common $O(1)$ scale, so that
$\lambda$ is interpretable and stable across corpora of differing speaking rate.
The indicator withholds latency credit from rollouts whose match rate falls
below $m_0$: as noted in Sec.~\ref{sec:awed}, delay is measured only on matched
words, so a rollout that emits little or emits poorly would otherwise be
rewarded for appearing fast. The gate is inspired by the quality-thresholded
reward of HPO~\cite{ouyang2026hierarchical}.

Rewards are converted to group-relative advantages,
$\hat{A}_i = (r_i - \mathrm{mean}(\mathbf{r})) / \mathrm{std}(\mathbf{r})$,
which removes the need for a learned critic, and the policy is updated with the
clipped GRPO objective
\begin{equation}
\begin{split}
  \mathcal{J}(\theta) = \mathbb{E}\Bigl[
    \tfrac{1}{G}\textstyle\sum_{i}\tfrac{1}{|o_i|}\textstyle\sum_{t}
    \min\bigl(&\rho_{i,t}\hat{A}_i,\;
    \mathrm{clip}(\rho_{i,t}, 1{-}\varepsilon, 1{+}\varepsilon)\hat{A}_i\bigr)
  \Bigr] \\
  - \;\beta\, \mathbb{D}_{\mathrm{KL}}&\bigl[\pi_\theta \,\|\, \pi_{\mathrm{ref}}\bigr],
\end{split}
\label{eq:grpo}
\end{equation}
with $\rho_{i,t} = \pi_\theta(o_{i,t} \mid s_{i,t}) / \pi_{\theta_{\mathrm{old}}}(o_{i,t} \mid s_{i,t})$.

We train at a single conditioning delay. Nothing in~\eqref{eq:reward} references
$\tau$, so the policy is not rewarded for waiting a prescribed number of frames
but for committing once the evidence suffices; since the delay condition is
shared across operating points through the same AdaRMSNorm pathway, the learned
behaviour is expected to transfer across the range. Sec.~\ref{sec:results}
confirms that it does.
\section{Experimental Setup}

\subsection{Data Preparation}
This work uses English ASR data exclusively. We aggregate twelve publicly available corpora: YODAS~\cite{li2023yodas}, LibriHeavy~\cite{kang2024libriheavy}, VoxPopuli~\cite{lugosch2022pseudo}, GigaSpeech~\cite{chen2021gigaspeech}, Fisher~\cite{cieri2004fisher}, Common Voice~\cite{ardila2020common}, LibriSpeech~\cite{panayotov2015librispeech}, National Speech Corpus~\cite{koh2019building}, Speech Accessibility Project~\cite{hasegawa2024community}, Switchboard~\cite{godfrey1992switchboard}, Europarl~\cite{koehn2005europarl}, and CALLHOME~\cite{canavan1997callhome}, amounting to approximately 175k hours of raw audio. Only utterances between 1\,s and 30\,s in duration are retained for training. Word-level timestamps, which our training objective requires, are obtained by forced alignment using Qwen3-ForceAligner~\cite{shi2026qwen3}.

\textcolor{black}{Evaluation follows the HuggingFace Open ASR Leaderboard~\cite{srivastav2025open}, on its eight
English test sets: AMI, Earnings22, GigaSpeech, LibriSpeech test-clean,
LibriSpeech test-other, SPGISpeech, TED-LIUM, and VoxPopuli. Together these
cover read speech, meetings, earnings calls, parliamentary and lecture speech, so that no single domain dominates the aggregate. For each set we compute WER by pooling errors and reference words over all of its utterances, and report the unweighted mean of the eight resulting values; the same hypotheses are used to compute AWED as defined in Sec.~\ref{sec:awed}. Text normalization is performed with an in-house normalizer rather than the leaderboard default, and is applied identically to every system compared in this paper, so all reported numbers are internally consistent but not directly comparable to published leaderboard
entries.}

\subsection{Implementation Details}
Our experiments build upon the Voxtral-Mini-4B-Realtime\footnote{\url{https://huggingface.co/mistralai/Voxtral-Mini-4B-Realtime-2602}} checkpoint using its Hugging Face implementation. Only the language decoder is updated via LoRA ($r=\alpha=256$), while the causal audio encoder remains frozen. The training procedure comprises two sequential stages. Stage~1 performs supervised fine-tuning for one epoch. Following Voxtral-Realtime~\cite{liu2026voxtral}, the structural streaming delay $\tau$ is uniformly sampled from 1 to 30 frames ($80$--$2400\,\mathrm{ms}$, where $1\text{ frame} = 80\,\mathrm{ms}$) to cover a broad spectrum of lookahead configurations. To prevent premature truncation, six padding tokens are appended post-transcript, positioning the EOS token 6 frames ($480\,\mathrm{ms}$) beyond the reference ending. Stage~2 applies GRPO to the Stage~1 checkpoint with $\tau$ fixed at 6 frames, the optimal operating point identified in~\cite{liu2026voxtral}.

\section{Results and analysis}
\label{sec:results}

\subsection{Main Results}
\label{sec:main_results}

\begin{table*}[t]
\centering
\caption{Performance across structural delay settings ($\tau \in \{1, 3, 6, 12\}$ frames) over eight benchmarks. Latency metrics ($p_{50}, p_{90}, p_{95}$) are reported in seconds via AWED; WER is reported as mean Micro-WER (\%).}
\label{tab:main_results}
\small
\setlength{\tabcolsep}{3.5pt} 
\begin{tabular}{l cccc c cccc c cccc c cccc}
\toprule
 & \multicolumn{4}{c}{\textbf{Delay 80\,ms ($\tau=1$)}} & & \multicolumn{4}{c}{\textbf{Delay 240\,ms ($\tau=3$)}} & & \multicolumn{4}{c}{\textbf{Delay 480\,ms ($\tau=6$)}} & & \multicolumn{4}{c}{\textbf{Delay 960\,ms ($\tau=12$)}} \\
\cmidrule{2-5} \cmidrule{7-10} \cmidrule{12-15} \cmidrule{17-20}
\textbf{Model} & \textbf{WER} & $p_{50}$ & $p_{90}$ & $p_{95}$ & & \textbf{WER} & $p_{50}$ & $p_{90}$ & $p_{95}$ & & \textbf{WER} & $p_{50}$ & $p_{90}$ & $p_{95}$ & & \textbf{WER} & $p_{50}$ & $p_{90}$ & $p_{95}$ \\
\midrule
Exp 0 (Voxtral Realtime)   & 12.67 & 0.77 & 1.06 & 1.32 & & 9.57 & 0.93 & 1.10 & 1.23 & & 8.31 & 1.17 & 1.33 & 1.41 & & 7.77 & 1.65 & 1.80 & 1.87 \\
Exp 1d (Stage 1 SFT)       & 10.91 & 0.75 & \textbf{0.89} & \textbf{0.95} & & 9.14 & 0.91 & 1.04 & 1.11 & & 8.24 & 1.15 & 1.28 & 1.35 & & 7.75 & 1.63 & 1.76 & 1.85 \\
Exp 4c (Stage 2 GRPO) & \textbf{8.76} & \textbf{0.74} & \textbf{0.89} & \textbf{0.95} & & \textbf{8.18} & \textbf{0.86} & \textbf{1.00} & \textbf{1.07} & & \textbf{7.84} & \textbf{1.04} & \textbf{1.18} & \textbf{1.26} & & \textbf{7.66} & \textbf{1.53} & \textbf{1.69} & \textbf{1.76} \\
\bottomrule
\end{tabular}
\end{table*}

\begin{figure}[htbp!]
    \centering
    \makebox[\linewidth][c]{%
        \includegraphics[width=1.0\linewidth]{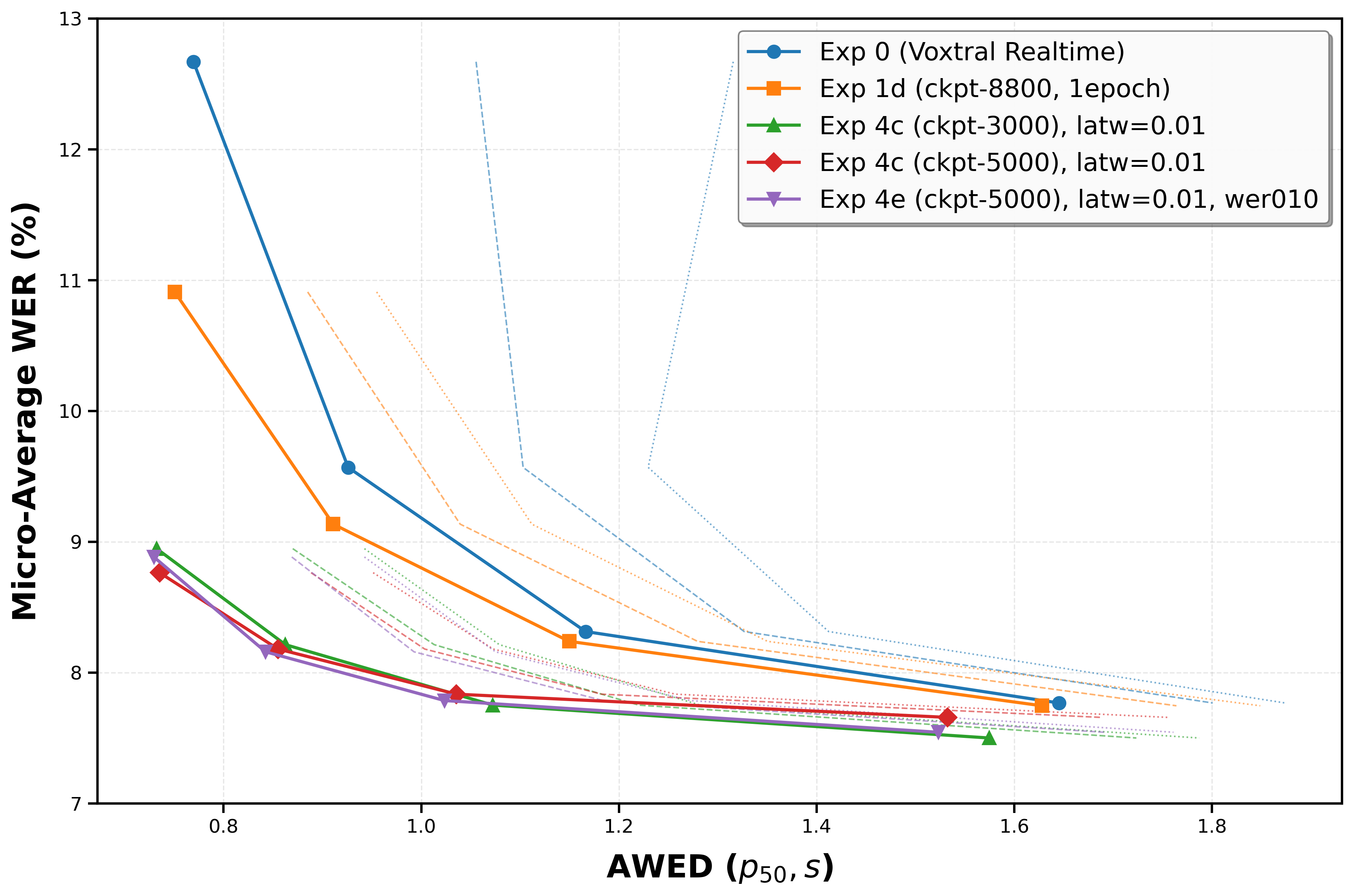}%
    }
    \caption{Accuracy-latency Pareto frontiers. Y-axis: mean Micro-WER (\%); X-axis: output delay (AWED $p_{50}$ in seconds). Dashed lines indicate $p_{90}$ and $p_{95}$ long-tail delays. Exp~4c (Stage~2 GRPO) consistently dominates the frontiers.}
    \label{fig:pareto_frontier_overall}
\end{figure}

To evaluate the accuracy-latency trade-off, Table~\ref{tab:main_results} and Fig.~\ref{fig:pareto_frontier_overall} present the quantitative performance and corresponding Pareto frontiers across four structural delay settings ($\tau \in \{1, 3, 6, 12\}$ frames, where $1\text{ frame} = 80\,\text{ms}$). Here, recognition accuracy is reported as the mean Micro-WER across eight inference benchmarks, while latency is measured via AWED ($p_{50}$ serves as the primary metric, with $p_{90}$ and $p_{95}$ representing long-tail delays as indicated by dashed lines in the Pareto plots).

Compared to the pretrained Voxtral-Realtime backbone (Exp~0), Stage~1 SFT (Exp~1d) achieves comparable overall performance while offering pronounced gains under tight delay constraints. At an $80\,\text{ms}$ structural delay ($\tau=1$), Stage~1 SFT lowers WER from $12.67\%$ to $10.91\%$ ($13.9\%$ relative reduction) at similar latency $\text{AWED } p_{50}$. Moreover, it substantially mitigates long-tail emission lag, reducing $p_{90}$ from $1.06\,\text{s}$ to $0.89\,\text{s}$ and $p_{95}$ from $1.32\,\text{s}$ to $0.95\,\text{s}$, which stabilizes token emissions and provides a well-behaved checkpoint for Stage~2.

Building upon the Stage~1 checkpoint, Stage~2 GRPO post-training—using our grounded hyperparameter choices of $\lambda = 0.01$ and $5\text{k}$ steps (Exp~4c, Section~\ref{sec:ablation})—yields substantial performance gains across the entire latency spectrum. Under the default setting ($480\,\text{ms}$ delay budget), Exp~4c brings WER down to $7.84\%$ (vs. $8.31\%$ in Exp~0) while reducing AWED $p_{50}$ latency to $1.04\,\text{s}$ (down from $1.17\,\text{s}$). Crucially, despite being conditioned on a fixed $\tau = 6$ during training, these optimizations generalize effectively across the entire latency spectrum. In particular, under the $80\,\text{ms}$ setting ($\tau = 1$), Exp~4c pushes WER down to $8.76\%$—achieving relative error reductions of $19.7\%$ and $30.8\%$ relative to Exp~1d and Exp~0, respectively—while maintaining an AWED $p_{50}$ latency comparable to Exp~1d.

\subsection{Ablation Study}
\label{sec:ablation}

\begin{figure}[htbp!]
    \centering
    
    \begin{subfigure}[b]{\linewidth}
        \centering
        \includegraphics[trim=2mm 2mm 2mm 2mm, clip, width=\linewidth]{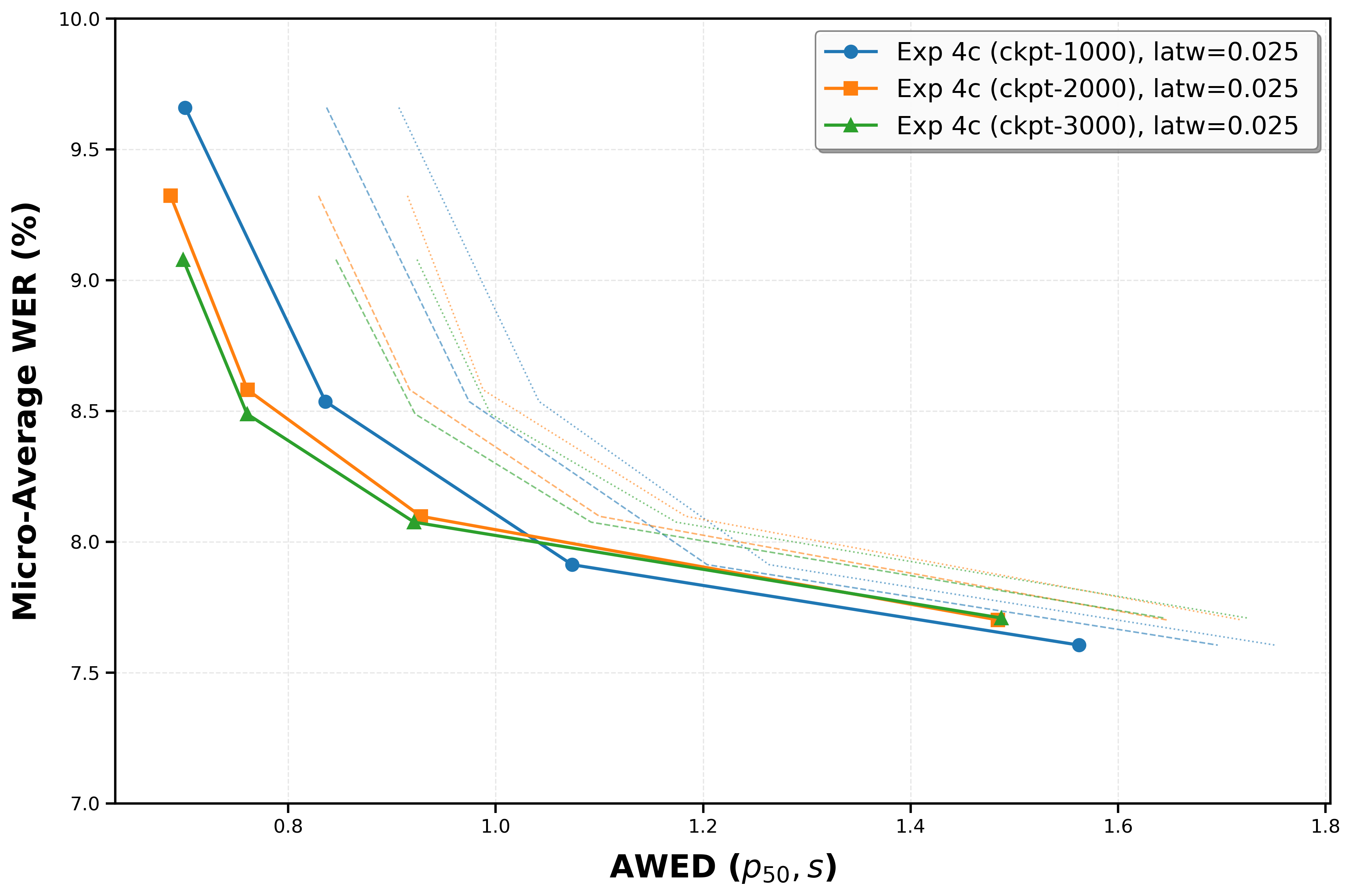}
        \caption{Impact of GRPO training steps ($\lambda = 0.025$).}
        \label{fig:pareto_steps}
    \end{subfigure}
    
    \vspace{2mm} 

    \begin{subfigure}[b]{\linewidth}
        \centering
        \includegraphics[trim=2mm 2mm 2mm 2mm, clip, width=\linewidth]{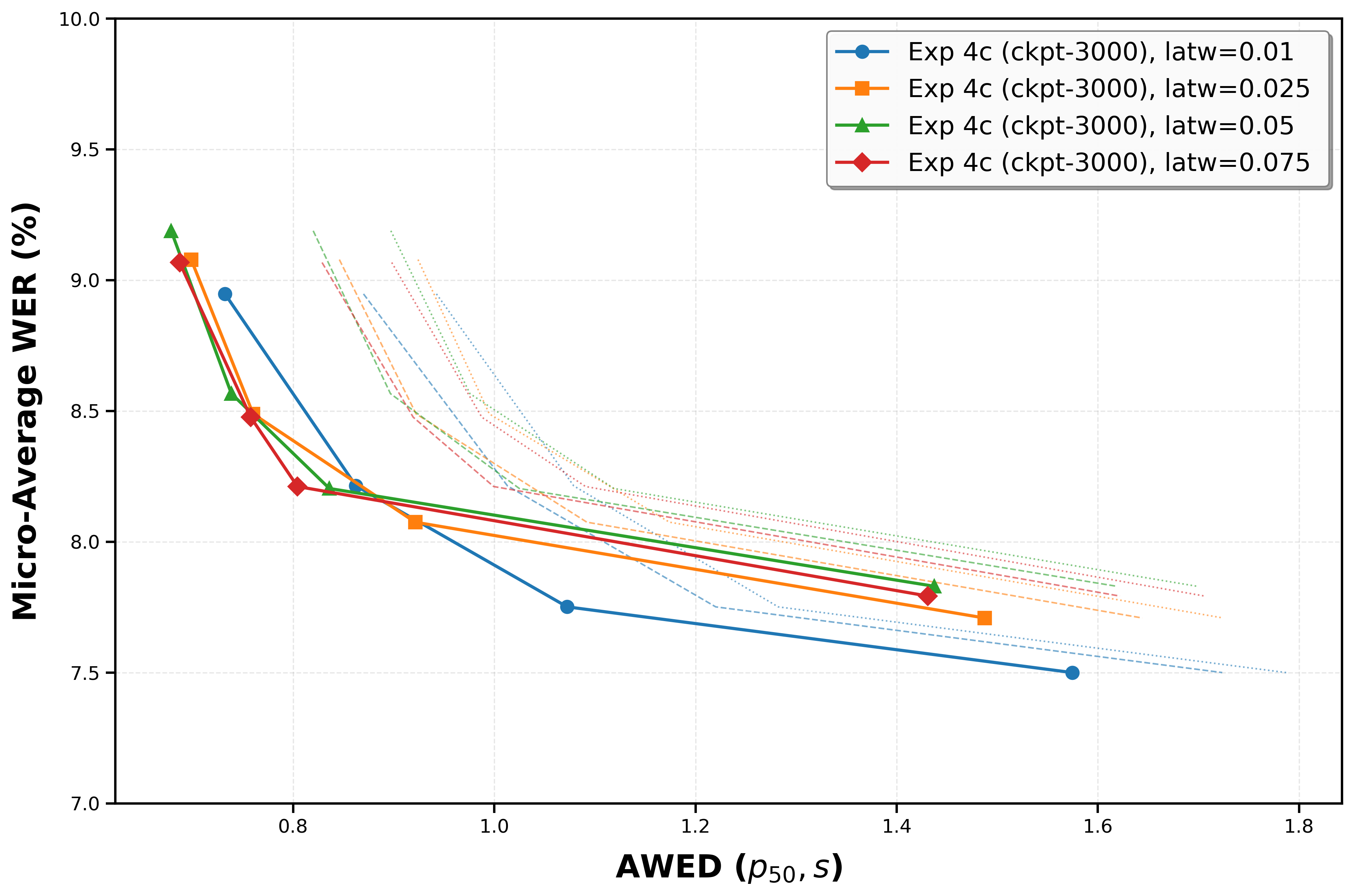}
        \caption{Impact of latency reward weight $\lambda$.}
        \label{fig:pareto_weights}
    \end{subfigure}

    \caption{Pareto frontiers illustrating the trade-off between recognition error (Micro-Average WER across 8 benchmarks) and streaming delay (AWED $p_{50}$). The ablations evaluate: (a) policy convergence over GRPO training steps, and (b) model sensitivity to the latency penalty coefficient $\lambda$.}
    \label{fig:pareto_frontier}
\end{figure}

This section presents systematic ablations over training duration and latency penalty weight to validate our optimal GRPO recipe ($\lambda = 0.01$, $5\text{k}$ steps).

First, regarding \textbf{training duration}, we evaluate checkpoints across $1\text{k}$ to $3\text{k}$ steps under $\lambda = 0.025$. The $3\text{k}$-step setup markedly outperforms $1\text{k}$ steps and achieves comparable trade-offs to $2\text{k}$ steps across the Pareto curve (\cref{fig:pareto_steps}). Extended training to $3\text{k}$ steps yields additional subtle WER gains at higher latency budgets, confirming that $3\text{k}$--$5\text{k}$ steps are sufficient for full convergence without over-fitting.

Second, concerning the \textbf{latency reward weight ($\lambda$)}, while higher penalties ($\lambda \in \{0.025, 0.05, 0.075\}$) yield near-identical trade-off profiles, a milder penalty ($\lambda = 0.01$) demonstrates distinct advantages (\cref{fig:pareto_weights}). Although it lags slightly at the tightest $80\,\text{ms}$ delay budget, $\lambda = 0.01$ unlocks noticeable accuracy gains as the delay budget expands. This indicates that a smaller weight avoids over-constraining the model's capacity, preserving higher accuracy for inference-time delay tuning.
\section{Conclusion}

\bibliographystyle{IEEEtran}
{\scriptsize\linespread{0.88}\selectfont\bibliography{refs}}

@article{shi2026qwen3,
  title={Qwen3-asr technical report},
  author={Shi, Xian and Wang, Xiong and Guo, Zhifang and Wang, Yongqi and Zhang, Pei and Zhang, Xinyu and Guo, Zishan and Hao, Hongkun and Xi, Yu and Yang, Baosong and others},
  journal={arXiv preprint arXiv:2601.21337},
  year={2026}
}

@inproceedings{li2023yodas,
  title={Yodas: Youtube-oriented dataset for audio and speech},
  author={Li, Xinjian and Takamichi, Shinnosuke and Saeki, Takaaki and Chen, William and Shiota, Sayaka and Watanabe, Shinji},
  booktitle={2023 IEEE Automatic Speech Recognition and Understanding Workshop (ASRU)},
  pages={1--8},
  year={2023},
  organization={IEEE}
}

@inproceedings{kang2024libriheavy,
  title={Libriheavy: A 50,000 hours ASR corpus with punctuation casing and context},
  author={Kang, Wei and Yang, Xiaoyu and Yao, Zengwei and Kuang, Fangjun and Yang, Yifan and Guo, Liyong and Lin, Long and Povey, Daniel},
  booktitle={ICASSP 2024-2024 IEEE International Conference on Acoustics, Speech and Signal Processing (ICASSP)},
  pages={10991--10995},
  year={2024},
  organization={IEEE}
}

@inproceedings{lugosch2022pseudo,
  title={Pseudo-labeling for massively multilingual speech recognition},
  author={Lugosch, Loren and Likhomanenko, Tatiana and Synnaeve, Gabriel and Collobert, Ronan},
  booktitle={ICASSP 2022-2022 IEEE International Conference on Acoustics, Speech and Signal Processing (ICASSP)},
  pages={7687--7691},
  year={2022},
  organization={IEEE}
}

@article{chen2021gigaspeech,
  title={Gigaspeech: An evolving, multi-domain asr corpus with 10,000 hours of transcribed audio},
  author={Chen, Guoguo and Chai, Shuzhou and Wang, Guanbo and Du, Jiayu and Zhang, Wei-Qiang and Weng, Chao and Su, Dan and Povey, Daniel and Trmal, Jan and Zhang, Junbo and others},
  journal={arXiv preprint arXiv:2106.06909},
  year={2021}
}

@inproceedings{cieri2004fisher,
  title={The Fisher corpus: A resource for the next generations of speech-to-text.},
  author={Cieri, Christopher and Miller, David and Walker, Kevin},
  booktitle={LREC},
  volume={4},
  pages={69--71},
  year={2004}
}

@inproceedings{ardila2020common,
  title={Common voice: A massively-multilingual speech corpus},
  author={Ardila, Rosana and Branson, Megan and Davis, Kelly and Kohler, Michael and Meyer, Josh and Henretty, Michael and Morais, Reuben and Saunders, Lindsay and Tyers, Francis and Weber, Gregor},
  booktitle={Proceedings of the twelfth language resources and evaluation conference},
  pages={4218--4222},
  year={2020}
}

@inproceedings{panayotov2015librispeech,
  title={Librispeech: an asr corpus based on public domain audio books},
  author={Panayotov, Vassil and Chen, Guoguo and Povey, Daniel and Khudanpur, Sanjeev},
  booktitle={2015 IEEE international conference on acoustics, speech and signal processing (ICASSP)},
  pages={5206--5210},
  year={2015},
  organization={IEEE}
}

@inproceedings{koh2019building,
  title={Building the singapore english national speech corpus},
  author={Koh, Jia Xin and Mislan, Aqilah and Khoo, Kevin and Ang, Brian and Ang, Wilson and Ng, Charmaine and Tan, Ying-Ying},
  booktitle={Proc. Interspeech 2019},
  pages={321--325},
  year={2019}
}

@article{hasegawa2024community,
  title={Community-supported shared infrastructure in support of speech accessibility},
  author={Hasegawa-Johnson, Mark and Zheng, Xiuwen and Kim, Heejin and Mendes, Clarion and Dickinson, Meg and Hege, Erik and Zwilling, Chris and Channell, Marie Moore and Mattie, Laura and Hodges, Heather and others},
  journal={Journal of Speech, Language, and Hearing Research},
  volume={67},
  number={11},
  pages={4162--4175},
  year={2024},
  publisher={American Speech-Language-Hearing Association}
}

@inproceedings{godfrey1992switchboard,
  title={SWITCHBOARD: Telephone speech corpus for research and development},
  author={Godfrey, John J and Holliman, Edward C and McDaniel, Jane},
  booktitle={[Proceedings] ICASSP-92: 1992 IEEE International Conference on Acoustics, Speech, and Signal Processing},
  volume={1},
  pages={517--520},
  year={1992},
  organization={IEEE}
}

@inproceedings{koehn2005europarl,
  title={Europarl: A parallel corpus for statistical machine translation},
  author={Koehn, Philipp},
  booktitle={Proceedings of machine translation summit x: papers},
  pages={79--86},
  year={2005}
}

@misc{canavan1997callhome,
  author       = {Canavan, Alexandra and Graff, David and Zipperlen, George},
  title        = {{CALLHOME} {A}merican {E}nglish Speech {LDC97S42}},
  howpublished = {Web Download},
  publisher    = {Linguistic Data Consortium},
  address      = {Philadelphia, PA, USA},
  year         = {1997},
  doi          = {10.35111/exq3-x930}
}

@article{liu2026voxtral,
  title={Voxtral realtime},
  author={Liu, Alexander H and Ehrenberg, Andy and Lo, Andy and Sun, Chen-Yo and Lample, Guillaume and Delignon, Jean-Malo and Chandu, Khyathi Raghavi and von Platen, Patrick and Muddireddy, Pavankumar Reddy and Arora, Rohin and others},
  journal={arXiv preprint arXiv:2602.11298},
  year={2026}
}

@article{srivastav2025open,
  title={Open ASR leaderboard: Towards reproducible and transparent multilingual and long-form speech recognition evaluation},
  author={Srivastav, Vaibhav and Zheng, Steven and Bezzam, Eric and Bihan, Eustache Le and Koluguri, Nithin Rao and {\.Z}elasko, Piotr and Majumdar, Somshubra and Moumen, Adel and Gandhi, Sanchit},
  journal={arXiv preprint arXiv:2510.06961},
  year={2025}
}

@article{shao2024deepseekmath,
  title={DeepSeekMath: Pushing the Limits of Mathematical Reasoning in Open Language Models},
  author={Shao, Zhihong and Wang, Peiyi and Zhu, Qihao and Xu, Runxin and Song, Junxiao and Bi, Xiao and Zhang, Haowei and Zhang, Mingchuan and Li, Y. K. and Wu, Y. and Guo, Daya},
  journal={arXiv preprint arXiv:2402.03300},
  year={2024}
}

@article{shivakumar2025grpo,
  title={Group Relative Policy Optimization for Speech Recognition},
  author={Shivakumar, Prashanth Gurunath and Gu, Yile and Gandhe, Ankur and Bulyko, Ivan},
  journal={arXiv preprint arXiv:2509.01939},
  year={2025}
}

@article{zeghidour2025streaming,
  title={Streaming sequence-to-sequence learning with delayed streams modeling},
  author={Zeghidour, Neil and Kharitonov, Eugene and Orsini, Manu and Volhejn, V{\'a}clav and de Marmiesse, Gabriel and Grave, Edouard and P{\'e}rez, Patrick and Mazar{\'e}, Laurent and D{\'e}fossez, Alexandre},
  journal={arXiv preprint arXiv:2509.08753},
  year={2025}
}

@inproceedings{ouyang2026hierarchical,
  title={Hierarchical Policy Optimization for Simultaneous Translation of Unbounded Speech},
  author={Ouyang, Siqi and Ding, Shuoyang and Hrinchuk, Oleksii and Lavrukhin, Vitaly and Yan, Brian and Ginsburg, Boris and Li, Lei},
  booktitle={Proceedings of the 64th Annual Meeting of the Association for Computational Linguistics (Volume 1: Long Papers)},
  pages={1772--1787},
  year={2026},
  publisher={Association for Computational Linguistics}
}

\end{document}